\documentclass[a4paper,12pt]{article}
\usepackage{amsfonts}

\newcommand{\cleqn}{\setcounter{equation}{0}}
\newcommand{\clth}{\setcounter{thm}{0}}
\newcommand {\sectionnew}[1]{\section{#1}\cleqn\clth}
\newcommand{\beq}{\begin{equation}}
\newcommand{\eeq}{\end{equation}}
\newcommand{\beqa}{\begin{eqnarray}}
\newcommand{\eeqa}{\end{eqnarray}}
\newcommand{\nn}{\nonumber \\}
\newtheorem{thm}{Theorem}

\newtheorem{prop}[thm]{Proposition}
\newcommand {\np}[1]{ {\textrm{:}}{#1}{\textrm{:}} } 
\newcommand {\boldsymbol}[1]{ {\underline{{#1}}} }

\def \s {\sigma}
\def \r {\rho}
\def \e {\mathrm{\bf e}}
\def \ex {\mathrm{e}}
\def \ep {\varepsilon}
\def \h {\frac{1}{2}}
\def \z {\zeta}
\def \W {W_{1+\infty}}
\def \A {{\mathcal A} }
\def \g {{\mathcal G} }
\def \d {\delta}
\def \D {\Delta}
\def \o {\omega}
\def \la {\langle}
\def \ra {\rangle}
\def \H {{\mathcal H}}
\def \Zset {{\mathbb Z}}
\def \Nset {{\mathbb N}}
\def \G   {\Gamma}
\def \q   {{\mathrm {\bf q}}}
\def \L   {\Lambda}
\def \l   {\lambda}
\begin{document}
\title{ RCFT extensions of $\W$ in terms of \\
bilocal fields}
\author{ L.S. Georgiev
\thanks{E-mail: lgeorg@inrne.acad.bg}
\quad I.T. Todorov
\thanks{E-mail: todorov@inrne.acad.bg}
\\
\normalsize\textit{ Institute for Nuclear
Research and Nuclear Energy}\\
\normalsize\textit{ Tsarigradsko Chaussee 72,
BG-1784 Sofia, Bulgaria }
\\
}
\date{}
\maketitle

\begin{abstract}
The rational conformal field theory (RCFT) extensions of
$\W$ at $c=1$ are in one-to-one correspondence with
1-dimensional integral lattices $L(m)$. Each extension
is associated with a pair of oppositely charged ``vertex
operators" of charge square $m\in\Nset$. Their product
defines a bilocal field $V_m(z_1,z_2)$ whose expansion
in powers of $z_{12}=z_1-z_2$ gives rise to a series of
(neutral) local quasiprimary fields $V^l(z,m)$ (of dimension
$l+1$). The associated bilocal exponential of a normalized
current generates the $\W$ algebra spanned by the $V^l(z,1)$
(and the unit operator). The extension of this
construction to higher (integer) values of the central
charge $c$ is also considered.

Applications to a quantum
Hall system require computing characters (i.e., chiral
partition functions) depending not just on the modular
parameter $\tau$, but also on a chemical potential $\z$.
We compute such a $\z$ dependence of orbifold characters,
thus extending the range of applications of a recent study
of affine orbifolds.

\end{abstract}
\vspace{-18cm}
\begin{flushright}
{\tt{ hep-th/9710134}}\\
{\tt{ INRNE-TH-97/11}}
\end{flushright}
\newpage
\section{Introduction}
The unitary positive energy representations of $\W$ \cite{kr1}
correspond to positive integer central charges $c\in\Nset$.
A standard tool in their study \cite{kr1,kr2} \cite{fkrw}
\cite{afmo} is the embedding of all irreducible modules for
a given $c$ into the Fock space of $c$ pairs of free Weyl
fermion fields $\psi^*_i(z)$.
It has been used in \cite{bgt} to develop a conformal
(quantum) field theory (CFT) approach to $\W$ based on the
expansion of the bilocal field
\beq\label{1.1}
V_1(z_1,z_2)=\sum_{i=1}^c \Bigl(\psi^*_i(z_1)\psi_i(z_2)-
\frac{1}{z_{12}}\Bigr),\;\;\;  z_{12}=z_1-z_2
\eeq
into local quasiprimary fields. $V_1$ is invariant under
SO(2$c$)($\supset \mathrm{U(1)}^{\times c}$) gauge
transformations in
the $2c$ -dimensional space of $\psi_i$ and $\psi^*_i$
(it is, in particular, $\mathrm{U(1)}^{\times c}$ neutral).
The index $1$ on $V$ indicates the charge square of $\psi_i$
(and $\psi^*_i$) which coincides with the (total) conformal
dimension of $V_1$: there is a representation $U(\r)$ of the
multiplicative group of positive reals ($\r>0$) in the state
space of our CFT such that
\beq\label{1.2}
U(\r)V_1(z_1,z_2){U(\r)}^{-1} = \r V_1(\r z_1,\r z_2) \;\;\;
(\mathrm{and} \; U(\r)|0\ra =|0\ra )
\eeq
($|0\ra$ representing the unique zero-energy state,
the {\em vacuum} of the theory).
The chiral algebra $W_c$ of $c$ Weyl fermions
(and their conjugates) corresponds to the simplest orthogonal
lattice, $\Zset^c$, of admissible charges: $W_c=\A(\Zset^c)$.
In general, the abelian compact Lie group
$\mathrm{U(1)}^{\times c}$ defines a lattice in
$\mathrm{u(1)}^{\otimes c}$ of elements $\o$ satisfying
$\ex^{2\pi i \o} =1$ (where 1 stands for the group unit).
Equipped with a positive definite integral quadratic form,
$(\o|\o') \in\Zset \;\; |\o|^2=(\o|\o) \in \Nset$ for
$\o \neq 0$, $L$ gives rise
to a chiral algebra $\A(L)$ whose positive energy (localizable)
representations give rise to an RCFT (see, e.g., Sec. 1 of
\cite{kt}). Singling out  a current $H(z)$ in the Kac-Moody
algebra $\mathrm{\hat{u}(1)}^{\otimes c}$ we can view $\A(L)$
as an RCFT extension of the level $c$ $\W$ algebra which
contains $H$.
We define a bilocal field $V_m(z_1,z_2)$, expressed as a bilocal
exponential in $H$ - see (\ref{2.14}) (generalizing (\ref{1.1})
to any positive integer charge square $m$) and establish for it
an {\em operator product expansion} (OPE) of quasiprimary fields
$V^l(z,m)$ of dimension $l+1 \ (l=0,1,\ldots)$. We prove, however,
that only for $m=1$, in which case the fields $V^i(z)= V^l(z,1)$
are expressed in terms of an orthonormal set of u(1) currents
$J^i$,
\beq\label{1.3}
\la J^i(z_1)J^j(z_2) \ra =\d_{ij} z_{12}^{-2} \;\; i,j=1,\ldots ,c,
\eeq
do the $V^l$ (and the unit operator) close an infinite dimensional
Lie algebra (which coincides with the level $c$ $\W$).
We work out the details, establishing exact OPE formulae, in Sec. 2
for the case $c=1$ and indicate the extension to arbitrary
$c\in\Nset$ in Sec. 3.
\sectionnew{ Rational extensions of $\W$ for $c=1$. OPE of
quasiprimary bilocal fields}
\subsection{ Local field basis for $\W$ in terms of Schur
polynomials of a u(1) current and its derivatives}
The product of two oppositely charged fields $\phi^\pm(z)$
of charge square $m$ can be expressed as a normal ordered
exponential of a U(1) current \cite{dfz}:
\beq\label{2.1}
\phi^+(z_1)\phi^-(z_2)=z_{12}^{-m}
\np{\exp \Bigl\{\int\limits_{z_2}^{z_1} H(z)
\mathrm{d}z \Bigr\} },\;\;\;
z_{12}=z_1-z_2.
\eeq
(The subtractions involved  in the definition of normal
ordering can be formally described by dividing the exponential
by its vacuum expectation value.) The normalization of the
``Cartan current" $H$ can be read off the singular terms of the
OPE:
\beq\label{2.2}
H(z_1)H(z_2)\sim \frac{m}{z_{12}^m},\;\;\;
H(z_1)\phi^\pm(z_2)\sim \frac{\pm m}{z_{12}}\phi^\pm(z_2).
\eeq
For a $c=1$ theory the stress energy tensor $T(z)$ is expressed
in terms of $H$ by the Sugawara formula:
\beq\label{2.3}
T(z)=\frac{1}{2m} \np{ H^2(z)} \;\;\;
(\np{H(z_1)H(z_2)}\equiv H(z_1)H(z_2)-\frac{m}{z_{12}^2}).
\eeq
Its Laurent modes $L_n$ generate the Virasoro algebra:
\beq\label{2.4}
T(z)=\sum_{n \in \Zset} L_n z^{-n-2} \;\;\;
([L_n,L_{n'}]=(n-n')L_{n+n'}+\frac{c}{12}(n^3-n)\d_{n,-n'},\;\;
c=1).
\eeq

A central result in \cite{bgt} says that the bilocal field
(\ref{2.1}) is, for $m=1$, a generating functional for a
local field basis of the $\W$ algebra. In fact, it then
coincides (up to a c-number term) with (\ref{1.1}) for
central charge $c=1$. We shall reformulate this result
using the U(1) current alone ( without reference to the
free Weyl field  $\psi$) thus extending it to the more
general current algebra framework.

Denote the normalized U(1) current $\frac{1}{\sqrt{m}} H(z)$
by $J^0(z)$. Using the Taylor expansion of the integral
in the exponential in (\ref{2.1}),
\beq\label{2.5}
\int\limits_{z_2}^{z_1} J^0(z) \mathrm{d}z = \sum_{l=0}^\infty
\frac{z_{12}^{l+1}}{(l+1)!} \partial_2^l J^0(z_2) \;\;\
(\partial_2=\frac{\partial}{\partial z_2}; \;\;
\la J^0(z_1)J^0(z_2) \ra_0=z_{12}^{-2})
\eeq
we introduce the {\em Schur polynomials} (in the terminology
of \cite{kr3} and \cite{bhy}) of $J^0$ and its derivatives,
\beq\label{2.51}
J^l(z)=l! \ \np{S_{l+1}(J^0(z),\frac{\partial J^0(z)}{2!},\ldots,
\frac{\partial^l J^0(z)}{(l+1)!})},
\eeq
 setting
\beq\label{2.6}
V_1(z_1,z_2)=\frac{1}{z_{12}} \Bigl( \np{\exp\Bigl\{
\int\limits_{z_2}^{z_1} J^0(z) \mathrm{d}z \Bigr\}} -1 \Bigr)=
\sum_{l=0}^\infty \frac{z_{12}^l}{l!} J^l(z_2).
\eeq
In general, a Schur polynomial of $N$ variables is defined by
\beq\label{2.61}
S_N(t_1,t_2,\ldots, t_k,\ldots)= \sum_{
\begin{array}{c}
n_1,n_2,\ldots, n_k,\ldots \\
\sum\limits_{k=1}^\infty  k\ n_k=N
\end{array}
}
\frac{t_1^{n_1}t_2^{n_2}\cdots t_k^{n_k}\cdots}
{n_1! n_2!\cdots n_k! \cdots} \; .
\eeq
The series of all Schur polynomials (of an increasing number
of variables) is generated by
\beq\label{2.62}
\exp \Bigl(\sum_{k=1}^\infty t_k z^k\Bigr)=\sum_{N=0}^\infty z^N
S_N(t_1,t_2,\ldots, t_k,\ldots).
\eeq
$J^l(z)$, as computed from (\ref{2.5}) (\ref{2.6}), satisfy
the recurrence relation
\beq\label{2.7}
(l+1)J^l(z)=(J^0+\partial)l J^{l-1}=\cdots=
(J^0+\partial)^l J^0 \;
(\equiv (l+1)! S_{l+1} (J^0,..),\
\partial \equiv \frac{\mathrm{d}}{\mathrm{d}z});
\eeq
we have: $2J^1=(J^0)^2+\partial J^0,\;
3J^2=(J^0)^3 +3J^0\partial J^0 +\partial^2 J^0,\;
4J^3=(J^0)^4 +6(J^0)^2\partial J^0 +3(\partial J^0)^2+
4J^0 \partial^2 J^0 +\partial^3 J^0$ etc.
(Viewed as quantum operators these expressions should be
understood as normal products.) The local fields $J^l$
(which, together with the unit operator span the infinite
dimensional Lie algebra $\W$) are singled for their simple
commutation relations (\cite{bgt} Lemma 1.2) which are
coded in the (singular terms of the ) OPE
\beqa\label{2.8}
J^k(z_1)J^l(z_2) &\simeq& \sum_{\nu \geq 1}
\Bigl\{ \frac{l!}{(l-\nu)!}J^{k+l-\nu}(z_2)-
 (-1)^\nu \frac{k!}{(k-\nu)!}J^{k+l-\nu}(z_1)\Bigr\}
z_{12}^{-\nu-1} + \nn
&+&(-1)^k l! k! z_{12}^{-k-l-2}.
\eeqa
As we shall see  shortly, the bilocal field (\ref{2.6})
can also be expanded in a basis of local {\em quasiprimary}
fields $V^l(z)$ of dimension $l+1$ which are expressed
in terms of $J^l$ (\cite{bgt} Lemma 1.3) by
\beq\label{2.9}
V^l(z)=\frac{(l!)^2}{(2l)!}\sum_{k=0}^\infty (-1)^k
{l\choose k} {2l-k\choose l} \partial^k J^{l-k}(z).
\eeq
A characteristic feature of quasiprimary fields is
their orthogonality; indeed, the 2-point function of
a pair of $V^l$'s is
\beq\label{2.10}
\la V^l(z_1)V^{l'}(z_2)\ra_0= \frac{(l!)^2}{(2l)!}
z_{12}^{-2l-2} \d_{l,l'}
\eeq
($\la\cdots\ra_0$ being a shorthand for the vacuum expectation
value  $\la 0|\cdots|0\ra_0$).

According to \cite{kr1} each positive energy unitary irreducible
representation of $\W$ for $c=1$ is characterized by a single
real number $r$, the $J^0$ charge, and the existence of a unique
ground state $|r\ra$ such that
\beq\label{2.11}
(J_0^0-r)|r\ra=0=(L_0-\h r^2)|r\ra,\;
J_n^0|r\ra=0 (=L_n|r\ra)\; \mathrm{for} \; n>0.
\eeq
Here $J^0_n$ are the $J^0(z)$ Laurent modes, such that
\beq\label{2.12}
J^0(z)=\sum_{n\in\Zset} J^0_n z^{-n-1},\;\;
[J^0_n,J^0_{n'}]=n \d_{n,-n'}
\eeq
The {\em minimal quantum field theory} whose state space
includes the irreducible $\W$ modules $\H$ and $\H_r$
(the {\em vacuum module} and the charge $r\neq 0$ module
$\H_r$ ) is generated by a pair of oppositely charged fields
$\phi(z,\pm r)$. Its OPE algebra contains (the $\W$ currents and)
an infinite series of charged fields $\phi(z,n r)$, $n\in\Zset$
(relatively local to $J^0$ - see \cite{bmt}). For $r^2=m\in\Nset$
they are also local among themselves Bose or Fermi fields; more
precisely, they satisfy
\beqa\label{2.13}
z_{12}^{r_1 r_2} \phi(z_1, r_1)\phi(z_2,r_2)=
z_{21}^{r_1 r_2} \phi(z_2, r_2)\phi(z_1,r_1) \nn
\mathrm{for} \; r_{1,2} \in \Zset\sqrt{m} \;(m=1,2,\ldots
\mathrm{fixed}).
\eeqa
In this case we recover the algebra $\A_m$ generated by the fields
$\phi^\pm(z)=\phi(z,\pm \sqrt{m})$ satisfying (\ref{2.1}), which
provides an RCFT extension of $\W$. Moreover, as noted in
\cite{bgt} (sec. 3) any RCFT extension of $\W$ for $c=1$ is of
this type.

\subsection{ Expansion of the dipole field (\ref{2.1}) into
local quasiprimary fields $V^l(z,m)$}
We introduce for arbitrary $m \;(=1,2,\ldots) $ and $c=1$ the
counterpart of (\ref{1.1}), the {\em bilocal field}
\beq\label{2.14}
V_m(z_1,z_2)= z_{12}^{m-1} \phi^+(z_1)\phi^-(z_2) - z_{12}^{-1}=
\frac{1}{z_{12}} \Bigl( \np{\exp\Bigl\{
\int\limits_{z_2}^{z_1} H(z) \mathrm{d}z \Bigr\}} -1 \Bigr);
\eeq
it is {\em quasiprimary} of total dimension $1$ in the sense
that it satisfies
\beq\label{2.15}
[L_n, V_m(z_1,z_2)]=\sum_{i=1}^2 z_i^n
(z_i \frac{\partial}{\partial z_i} +\frac{n+1}{2})V_m(z_1,z_2)
\;\; \mathrm{for} \ n=0,\pm 1.
\eeq
(It is easily verified, in particular, that the differential
operators in the right hand side annihilate, for $n=0,\pm 1$,
the 2-point function $z_{12}^{-1}$.)
Eq. (\ref{2.14}) (with the current $H$ satisfying (\ref{2.2}))
provides a version of the vertex operator construction, thus
allowing to compute the correlation functions of $V_m$. We have,
in particular, $\la V_m(z_1,z_2)\ra_0=0$,
\beqa\label{2.16}
&&\la V_m(z_1,z_2)V_m(z_3,z_4)\ra_0=\frac{1}{z_{12}z_{34}}
\Bigl\{ (1-\eta)^{-m}-1 \Bigr\}=\nn
&&=\frac{m}{z_{13}z_{24}}\sum_{n=0}^\infty {m+n \choose m}
\frac{\eta^n}{n+1} ,\;\;\; \eta=\frac{z_{12}z_{34}}{z_{13}z_{24}};
\eeqa
\beqa\label{2.17}
&&\la V_m(z_1,z_2)V_m(z_3,z_4)V_m(z_5,z_6)\ra_0=
(z_{12}z_{34}z_{56})^{-m}
\Bigl\{ 2-\Bigl(\frac{z_{13}z_{24}}{z_{23}z_{14}}\Bigr)^m-\nn
&&-\Bigl(\frac{z_{35}z_{46}}{z_{36}z_{45}}\Bigr)^m-
\Bigl(\frac{z_{15}z_{26}}{z_{16}z_{25}}\Bigr)^m+
\Bigl(\frac{z_{13}z_{24}z_{35}z_{46}z_{15}z_{26}}
{z_{23}z_{14}z_{36}z_{45}z_{16}z_{25}}\Bigr)^m
\Bigr\}.
\eeqa
We shall describe the complete OPE of (\ref{2.14}) into local
Bose fields thus extending to arbitrary $m\in\Nset$ an earlier
result about $V_1$ (cf. Theorem 2.1 of \cite{bgt}).
\begin{thm}\label{t2.1}
The bilocal field (\ref{2.14}) admits an expansion in
integrals over local quasiprimary fields $V^l$
(of dimension $l+1$):
\beq\label{2.18}
V_m(z_1,z_2)=\sum_{l=0}^\infty \frac{z_{12}^l}{l!}
\int\limits_{z_2}^{z_1} p_l(z;z_1,z_2)V^l(z,m)\mathrm{d}z
\eeq
where $p_l$ is the (normalized) weight function
\beq\label{2.19}
p_l(z;z_1,z_2)=\frac{(2l+1)!}{(l!)^2 z_{12}^{2l+1}}
(z_1-z)^l (z-z_2)^l \;\;\;
\Bigl( \int\limits_{z_2}^{z_1} p_l(z;z_1,z_2)
\mathrm{d}z=1 \Bigr).
\eeq
The $2$-point functions of $V^l$ (which display their
orthogonality),
\beq\label{2.20}
\frac{l+1}{l!} \la V^l(z_1,m)V^{l'}(z_2,m)\ra_0=
C_{lm} z_{12}^{-2l-2} \d_{l,l'},
\eeq
involve the same normalization constant as the $3$-point function
\beq\label{2.21}
(l+1)\la V_m(z_1,z_2) V^l(z_3,m)\ra_0=C_{lm}
\frac{z_{12}^l}{(z_{13}z_{23})^{l+1}}.
\eeq
The (positive) constants $C_{lm}$ are uniquely determined by the
$4$-point function (\ref{2.16}). Conversely,$V^l(z,m)$ are
computed from $V_m(z_1,z_2)$ by differentiation:
\beq\label{2.22}
V^l(z,m)=\frac{l!}{(2l)!} \lim_{z_1,z_2 \rightarrow z}
\Bigl\{\partial_1^l (-\partial_2)^l
\Bigl( z_{12}^l V_m(z_1,z_2)\Bigr)\Bigr\}.
\eeq
Furthermore, $V^l$ are again related to the Schur polynomials
$S_n(H,\frac{\partial H}{2!},\ldots,\frac{\partial^{n-1} H}{n!})$
by (the counterpart of) Eq. (\ref{2.9}).
\end{thm}
The {\bf proof} follows, for the most part the argument in
\cite{bgt} and we shall
only sketch it emphasizing the new points.

The weight function (\ref{2.19}) satisfies the identity
\beq\label{2.23}
 \int\limits_{z_2}^{z_1} p_l(z;z_1,z_2) (z-z_3)^{-2l-2}
\mathrm{d}z=z_{13}^{-l-1}z_{23}^{-l-1}.
\eeq
Since, on the other hand, 2- and 3- point functions of
quasiprimary fields are determined (up to normalization
constants) by M\"{o}bius (conformal) invariance, this implies
\beq\label{2.24}
\la V_m(z_1,z_2) V^l(z_3,m) \ra_0 =\frac{z_{12}^l}{l!}
 \int\limits_{z_2}^{z_1} p_l(z;z_1,z_2)
\la V^l(z,m) V^l(z_3,m)\ra_0 \mathrm{d}z.
\eeq
Orthogonality of $V^l$ (reflected in (\ref{2.20}) ) is
a general property of quasiprimary fields of different
dimensions. Inserting the expansion (\ref{2.18}) for
$V_m(z_3,z_4)$ into (\ref{2.16}) and using (\ref{2.21})
we end up with a triangular set of linear equations for
$C_{lm}$:
\beq\label{2.25}
(1-\eta)^{-m}=1+\sum_{l=0}^\infty C_{lm}
\frac{\eta^{l+1}}{(l+1)!} F(l+1,l+1; 2l+2;\eta).
\eeq
Here we have used the following representation for the
Gauss hypergeometric function
\beq\label{2.26}
F(n,n;2n;\eta)=z_{13}^nz_{24}^n \int\limits_{z_2}^{z_1}
\frac{p_l(z;z_1,z_2)  \mathrm{d}z}{(z_1-z)^n(z_2-z)^n} \;\;
\Bigl( \eta=\frac{z_{12}z_{34}}{z_{13}z_{24}} \Bigr).
\eeq
We thus have
\beq\label{2.27}
\sum_{l=0}^\infty C_{lm} \frac{\eta^{l+1}}{(l+1)!}
\sum_{n=0}^\infty {l+n\choose n}^2 {2l+1+n\choose n}^{-1} \eta^n=
\sum_{l=0}^\infty {m+l\choose l+1} \eta^{l+1}
\eeq
The first seven $C_{lm}$ are captured  by the formula
\beq\label{2.28}
C_{lm}=m^{l+1}+{l+1\choose 4} \frac{m^{l-1}}{2l-1}
+{l+1\choose 6}\frac{5l^2-7l+6}{8(2l-1)(2l-3)} m^{l-3}
 \;\;
\mathrm{for} \; 0\leq l\leq 6.
\eeq
On the other hand, for $m=1$, we recover the expression
$C_{l1}=(l+1)! {2l\choose l}^{-1}$ of \cite{bgt}.
The important point is the existence of a unique solution
of the (infinite) system (\ref{2.27})  $C_{lm}>0$.

Finally, Lemma 1.3 of \cite{bgt} holds in this case, too,
and we can write
\beq\label{2.29}
V^l(z,m)=\frac{(l!)^2}{(2l)!}\sum_{k=0}^l
(-1)^k{l\choose k}{2l-k\choose l}
(l-k)!\partial^k S_{l-k+1}(H,\frac{\partial H}{2!},\ldots,
\frac{\partial^{l-k} H}{(l-k+1)!} )
\eeq
where $ S_{n+1}(H,\frac{\partial H}{2!},\ldots,
\frac{\partial^{n} H}{(n+1)!})=(H+\partial)^n H$
(cf. Eq. (\ref{2.7})) and we omit (here and in the following
remark) the normal product sign.

{\bf Remark 2.1.} In spite of the fact that $V^l(z)=V^l(z,1)$
is expressed by the same Schur polynomials as $V^l(z,m)$ -
only the argument $H$ is substituted by the normalized current
$J^0$ in $V^l(z,1)$ - the quasiprimary fields $V^l(z,m)$
do not close under commutation among themselves
( and the unit operator) for
$m\geq 2$. In effect the fields
$V^l(z,m)=m^{\frac{l+1}{2}}V^l(z)$ for $l=0,1,2$ generate
the whole $\W$ algebra spanned by $V^l(z)$ (and 1). On the
other hand, the fields $V^l(z,m)$ and $V^l(z)$ are linearly
independent for $l\geq3,\; m\geq 2$. Indeed, we have
\beqa\label{2.30}
&&V^3(z,m)=\frac{H^4(z)}{4} +\frac{1}{20} P_1(H,H',H'')=
\frac{m^2}{4} J^4(z)+\frac{m}{20}(2JJ''-3{J'}^2)(z) \nn
&&(\mathrm{for}\; H=\sqrt{m}J);\;
V^4(z,m)=\frac{1}{5} H^5+\frac{1}{7} H P_1, \; \mathrm{etc.}
\eeqa
Here $P_1$ is the first of a series of quasiprimary fields
that are quadratic forms $P_n$ of $H$ and its derivatives
(each term having dimension $2n+2$):
\beqa\label{2.31}
&&P_n(H,H',\ldots, H^{(2n)})=2\sum_{l=0}^{n-1}
(-1)^l a_l H^{(l)} H^{(2n-l)}+ (-1)^n a_n (H^{(n)})^2,\nn
&&a_l=\prod_{k=0}^{l-1} {2n-1+k\choose 2} \prod_{k=l}^{n-1}
{k+2\choose 2}.
\eeqa

\sectionnew{ Extension to arbitrary $c \in \Nset$. Discussion}
\subsection{ Additive bilocal fields and lattice
RCFT extensions}
The construction  of Sec. 2 extends to arbitrary positive
integer c due to the following simple observation.
\begin{prop}{\em (\cite{bgt}, Lemma 2.2)}
Let $V_i(z_1,z_2), 1\leq i\leq c$ be a commuting set of
bilocal fields with the properties of $V_1(z_1,z_2)$
{\em (\ref{2.6})}.
Then their sum generates the $\W$ algebra for central
charge $c$:
\beqa\label{3.1}
&&\sum_{i=1}^c V_i(z_1,z_2)=\sum_{l=0}^\infty \frac{z_{12}^l}{l!}
J^l(z_2) \Rightarrow \nn
&&[J^k(z_1),J^l(z_2)]=\sum_{\nu\geq 1} \Bigl\{(-1)^\nu {l \choose \nu}
J^{k+l-\nu}(z_2)-
{k \choose \nu}J^{k+l-\nu}(z_1) \}\d^{(\nu)}(z_{12})\Bigr\}-\nn
&&-(-1)^l\frac{k!l!c}{(k+l+1)!}\d^{(k+l+1)}(z_{12}).
\eeqa
If we use the bosonized expressions
\beq\label{3.2}
V_i(z_1,z_2)=\frac{1}{z_{12}} \Bigl( \np{\exp\Bigl\{
\int\limits_{z_2}^{z_1} J^0_i(z) \mathrm{d}z \Bigr\}} -1 \Bigr)
\;\;\mathrm{with}\;\;
[J^0_i(z_1),J^0_j(z_2)]=- \d'(z_{12})\d_{ij},
\eeq
then $J^l(z)=\sum_{i=1}^c J^l_i(z)$ where $J^l_i$ is expressed
as a Schur polynomial of $J^0_i(z)$ and its derivatives
\beq\label{3.3}
J^l_i(z)=l! \ S_{l+1}(J^0_i(z),\frac{\partial J^0_i(z)}{2!},\ldots,
\frac{\partial^l J^0_i(z)}{(l+1)!}).
\eeq
\end{prop}
Similarly, the basic quasiprimary fields $V^l(z)$ appear
as sums of mutually commuting $c=1$ fields $V^l_i(z,1)$ .

Let $W^{(c)}$ be the chiral algebra corresponding to the vacuum
representation of $\W$ of central charge $c$; it is spanned by
(normal ordered) polynomials of $V^l(z)$. The set of RCFT extensions
is much richer for $c>1$. It includes, in particular, local charged
fields' algebras $\A(L)$ associated with a $c$-dimensional integral
Euclidean charge lattice $L$. In other words, there is a basis
$( \q^1,\ldots,\q^c)$ in the space of $\mathrm{U(1)}^{\times c}$
charges such that the (symmetric) positive definite Gram matrix
$(\q^i|\q^j)$ of inner  products has integer entries. To each such
lattice corresponds a vertex algebra (for a recent review - see
\cite{k}) spanned by charged vertex operators $Y(\q,z)$ and
``Cartan currents" $H^{\q}(z)$ satisfying OPE of the type
\beq\label{3.4}
Y(\q,z_1)Y(-\q,z_2)=\frac{1}{z_{12}^{|\q|^2}}
\np{\exp \Bigl\{\int\limits_{z_2}^{z_1} H^{\q}(z)
\mathrm{d}z \Bigr\} } ,\;\;
H^{\q}(z_1)H^{\q'}(z_2)\sim \frac{(\q|\q')}{z_{12}^2},
\eeq
\beq\label{3.5}
Y(\q,z_1)Y(\q',z_2)\sim \ep (\q,\q') z_{12}^{(\q|\q')}
Y(\q+\q',z_2) \;\;(\mathrm{for} \; \q+\q' \neq 0)
\eeq
where $\ep$ is a $(\pm 1)$ valued cocycle that ensures the correct
spin-statistics relation between different vertex operators.
If the charge lattice is odd, -i.e., if there are vectors $\q\in L$
such that $|\q|^2 \equiv (\q|\q) $ is odd (which is the case  of
interest in applications to a quantum Hall system - see \cite{fst}
\cite{ctz} \cite{cz} and references therein) then the resulting
chiral algebra $\A(L)$ is a local $\Zset_2$ graded superalgebra.
In order to identify a $W^{(c)}$ subalgebra we should choose a
$1$-dimensional subspace in the $c$-dimensional space of currents,
thus singling out the current $J^0(z)=\sum_{i=1}^c J^0_i(z)$.
This is achieved, in particular, by selecting an ``electric charge"
vector ${\bf Q}$ in the dual lattice $L^*$ (whose square
$| {\bf Q}|^2$ gives the quantum Hall filling
fraction \cite{fst} \cite{ctz}).

The positive energy representations of the RCFT so obtained,
in which all vertex operators $Y(\q,z),\ \q\in L$ are single
valued, are in one-to-one correspondence with the elements of
the (finite) abelian group $L^*/ L$ of order
\beqa
|L^*/L| = \mathrm{det} \Bigl((\q^i|\q^j)\Bigr).\nonumber
\eeqa
Let $\G$ be any finite group of automorphisms of the chiral
algebra $\A(L)$ which leaves each element of its $W^{(c)}$
subalgebra invariant. (It is enough to demand the invariance
of the first three ``tensor currents" $V^l(z),\ l=0,1,2$ which
generate the entire $\W$ algebra.) There is a wealth of such
inner automorphism groups in the
case when $L$ is a semisimple Lie group lattice (for integer $c$
it may only
corresponds to a lattice of the A-D-E type, the basis ${\q^i}$ being
 identified with a basis of simple roots). Viewing $\G$ as a {\em
gauge group} we define the {\em observable algebra} as the subalgebra
 $\A(L)^\G$ of  $\G$   invariant elements of  $\A(L)$. The resulting
chiral algebra $\A(L)^\G$ provides another RCFT extensions of
 $W^{(c)} $ (see \cite {kt}).

The characters  of positive energy  irreducible representations of
a lattice current algebra span, in general, a finite dimensional
representation of the  subgroup of the modular group SL(2,$\Zset$)
generated by the matrices
\beqa\label{3.6}
 S=\left(\matrix{
0 &-1\cr
1 &0\cr}\right) \;  \mathrm{and} \;
T^2= \left(\matrix{
1 &2\cr
0 &1\cr}\right)
\eeqa
(If $L$ is an even lattice then they span a representation of the
full SL(2,$\Zset$).)
Recently, all weak modular invariants (i.e. invariants under
the group generated by (\ref{3.6}) have been classified -see
\cite{g} ).  On the other hand, given a {\em  non-exceptional
gauge group} of inner automorphisms $\G$ (such is, in particular,
any finite subgroup of U(n) or SU(n) ),it has been demonstrated in
\cite{kt} how to construct the specialized characters
(that only capture the energy distribution)
of the corresponding
$\A(L)^\G$ {\em orbifold modules} and to exhibit their modular
transformation properties.

{\bf Remark 3.1.} Albeit we thus arrive at a rich family of
manageable RCFT extensions of $\W$, the set of such extensions
is not exhaustive - even if we  admit outer automorphisms in
$\G$, for which no general construction of orbifold modules is known.
RCFT extensions of a different type are  obtained through the so
called conformal embedding (\cite {sw},\cite{bn},\cite
{ago},\cite{s},\cite{t},\cite{rst}). A simple example of this type
is provided by the level 5 $su(3)$-current algebra $\A_5(su(3))$ of
Virasoro central charge $c_k=\frac{8k}{3+k}=5$ for $k=5$, which can
be viewed as the {\em observable subalgebra} of the rank 5 lattice
chiral algebra $\A_1(su(6))$. Indeed, the vacuum space of
$\A_1(su(6))$ splits into a direct sum of two $\A_5(su(3))$ modules
(with the same central charge $c=5$, and hence, the same Virasoro
subalgebra): one with $su(3)$ weight $\L_0=(0,0)$ and ground state
energy $\D_0=0$ and a second one with $\L_1=(2,2)$ and a degenerate
27 dimensional ground state of energy (=conformal weight)
$\D_1={\frac{1}{24}}\Bigl\{\l_1(\l_1+3)+\l_2(\l_2+3)+\l_1\l_2 \Bigr\}
=1$ for $\l_1=\l_2=2$.Thus $\A_5(su(3))$ can be viewed as an RCFT
extension of either $W^{(5)}$ or the part of $W^{(6)}$ orthogonal
 to the $u(1)$ current $J^0$.It can not be viewed as the gauge
invariant part of the lattice algebra $\A_1(su(6))$ with respect to
a finite gauge group $\G$ since then the $\A_5(su(3))$ module of
highest  weight $\L_1$ should have an integer quantum dimension
(equal to the dimension of the associated irreducible representation
 of $\G$-see \cite{rst} and references therein) while it is, in fact,
irrational:
\beqa
d_q(\L)={\frac{[3]^2[6]}{[2]}}=(1+\sqrt{2})^2 \;\;\;
 \mathrm{for}\;\;\;
 q^8=-1.\nonumber
\eeqa
Only for $c=1$ we have listed all RCFT extensions of the $u(1)$
current algebra - cf.\cite{pt} -and hence all extensions of
$W^{(1)}$.
\subsection{ Computing character dependence on chemical
potential for orbifold models}
It is important, as far applications to the quantum Hall
effect are concerned, to compute the character dependence on
the chemical and electric potentials combined in a complex
variable $\z$ (see \cite{cz}; the $\z$ dependence carries
information about charge distribution and the so called
spectral flow). In the case of lattice orbifolds this
requires the following simple (but important) complement
of the algorithm presented in \cite{kt}.

To set the stage we reformulate briefly some notions
and results of \cite{kt} adapted to the present context.

Let $L$ be an integral euclidean lattice and let ${\bf Q}$
be a fixed vector in the dual lattice $L^*$, the ``electric
charge" (such a pair $(L,{\bf Q})$, satisfying some additional
requirements, is called a {\em chiral quantum Hall lattice}
in \cite{fst}). We shall assume, for the sake of convenience,
that $L$ is {\em maximal}, -i.e., there is no sublattice of
$L^*$ that is a proper extension of $L$. (This assumption
does not restrict the generality of the orbifold chiral
algebra we shall end up with.) Let $G$ be a connected compact
Lie group whose Lie algebra is generated by the 0-components
of all (dimension 1) currents in the chiral algebra $\A(L)$.
It can be hence assumed acting by unitary operators in each
irreducible (positive energy) $\A(L)$ module. $G$ is a rank
$l=\mathrm{dim}\ L$ Lie group: it has a maximal abelian
subgroup $\mathrm{U(1)}^{\times l}$ where $l=\mathrm{rank} \ L$;
the center $\Zset_{G} (\subset \mathrm{U(1)}^{\times l})$
of $G$ is nontrivial: it contains a (finite) subgroup
isomorphic to $|L^*/L|$. To any $\q\in L^*$ there corresponds
a pair of charged vertex operators $Y(\pm \q,z)$ (which
intertwine the vacuum representation of $\A(L)$ with the
charge $\pm \q$ representations) and hence a u(1) current
$H^{\q}(z)$ satisfying (\ref{3.4}) whose 0-component  $H^{\q}_0$
belongs to the Lie algebra $\g$ of $G$. ($\g$ may also involve
raising and lowering operators $E^{\pm \boldsymbol{\alpha}}$ for
${\boldsymbol{\alpha}} \in L$, $|\boldsymbol{\alpha}|^2=2$ whenever
$L$ contains such
vectors $\boldsymbol{\alpha}$.
Let $\G$ be a (non-exceptional) finite subgroup of $G$ acting
by inner automorphisms on $\A(L)$ and leaving the current
$H^{{\bf Q}}(z)$ invariant.
To each inner automorphism $\r\in \mathrm{Ad}\ \G$ of $\A(L)$
we make correspond a
$\boldsymbol{\beta} =\boldsymbol{\beta_{\r}} \in i\g$ such that
\beq\label{3.7}
\r(V)=b \ V \ b^{-1},\; b=\ex^{2\pi i \boldsymbol{\beta}},\;
 \G_{b}=\G_{\boldsymbol{\beta}}
\eeq
where $\G_b (\G_{\boldsymbol{\beta}} )$ is the centralizer
of $b(\boldsymbol{\beta})$ in
$\G$. (We identify vectors $\boldsymbol{\beta} \in L$ with
corresponding
Lie algebra element $H^{\boldsymbol{\beta}}_0$.)

Let $\H_{\boldsymbol{\l}},\ \boldsymbol{\l} \in L^*/L$ be
an irreducible positive
energy $\A(L)$ module. Define its character $\chi_{\boldsymbol{\l}}$
\beq\label{3.8}
\chi_{\boldsymbol{\l}}(\tau,\boldsymbol{\alpha})=
\mathrm{tr}_{ \H_{\boldsymbol{\l}}}
 (q^{L_0-\frac{c}{24}}
\ex^{2\pi i \boldsymbol{\alpha}}),\;
q=\ex^{2\pi i \tau},\; Im \ \tau >0;\; i\boldsymbol{\alpha} \in \g
\eeq
It follows that
\beq\label{3.9}
\chi_{\boldsymbol{\l}}(\tau,\boldsymbol{\alpha})=
\frac{1}{[\eta(\tau)]^l}
\sum_{\boldsymbol{\gamma}\in\ L + \boldsymbol{\l}}
q^{\frac{1}{2}(\boldsymbol{\gamma}|\boldsymbol{\gamma})}
\ex^{2\pi i (\boldsymbol{\gamma}|\boldsymbol{\alpha})}  \;\;\;\;
(\eta=q^{\frac{1}{24}}\prod_{n=1}^\infty (1-q^n), \
 l=\mathrm{rank} \ L).
\eeq
The character of a {\em $\boldsymbol{\beta}$-twisted module}
$\H_{\boldsymbol{\l}}^{\boldsymbol{\beta}}$
is (see Eq. (3.18) of \cite{kt})
\beq\label{3.10}
\chi_{\boldsymbol{\l}}^{\boldsymbol{\alpha},\boldsymbol{\beta}}
(\tau)=q^{ \h|\boldsymbol{\beta}|^2}
\chi_{\boldsymbol{\l}}(\tau, \boldsymbol{\alpha}-
\tau\boldsymbol{\beta}),\;\;
[\boldsymbol{\alpha},\boldsymbol{\beta}]=0.
\eeq
\begin{prop}
Let $\s$ be (the character of) an irreducible representation of
the centralizer $\G_{\boldsymbol{\beta}}(=\G_b)$ of
$\boldsymbol{\beta}$ in $\G$. Then the
projected module  $\H_{\boldsymbol{\l} \s}^{\boldsymbol{\beta}}$
is only nontrivial if
$\s$ and $\boldsymbol{\l}$ agree on $L^*/L$ and then the character
is
\beq\label{3.11}
\chi_{\boldsymbol{\l},\s}^{\boldsymbol{\beta}}(\tau,\z)=
\frac{1}{|\G_{\boldsymbol{\beta}}|}
\sum_{h=\ex^{2\pi i\boldsymbol{\alpha}}\in \G_{\boldsymbol{\beta}}}
\ex^{-2\pi i\z({\bf Q}|\boldsymbol{\beta})}
\chi_{\boldsymbol{\l}}^{\z{\bf Q}+\boldsymbol{\alpha},
\boldsymbol{\beta}}
(\tau)\s^*(h) \;
\eeq
$|\G_{\boldsymbol{\beta}}|$ standing for the number of elements of
$\G_{\boldsymbol{\beta}}$.
\end{prop}

Indeed, defining   $\H_{\boldsymbol{\l} \s}^{\boldsymbol{\beta}}$
  by
\beq\label{3.12}
\H_{\boldsymbol{\l} \s}^{\boldsymbol{\beta}}=P_\s
\H_{\boldsymbol{\l}}^{\boldsymbol{\beta}} \; \mathrm{where}\;
P_\s=\frac{\s(1)}{|\G_{\boldsymbol{\beta}}|}
\sum_{h\in\G_{\boldsymbol{\beta}}} \s^*(h)h \;
(=P_\s^2)
\eeq
one proves by a simple extension of the argument of Sec. 4.1 of
\cite{kt} that
\beq\label{3.13}
\chi_{\boldsymbol{\l},\s}^{\boldsymbol{\beta}}(\tau,\z)=
\mathrm{tr}_{ \H_{\boldsymbol{\l},\s}^{\boldsymbol{\beta}}}
(q^{L_0-H_0^{\boldsymbol{\beta}}+\h |\boldsymbol{\beta}|^2- \frac{c}{24}}
\ex^{2\pi i \z(H_0^{{\bf Q}}-({\bf Q}|\boldsymbol{\beta}))}).\;
\eeq
Instead of reproducing the details we shall work out an example of
physical interest (cf. \cite{cgt}).

{\bf Example 3.1.} Let $L$ be a 2-dimensional orthogonal lattice
with basis $\e^1,\e^2$ and Gram matrix
\beq\label{3.14}
\Bigl( (\e^i|\e^j) \Bigr)=
 \left(\matrix{
(\e^1|\e^1) &(\e^1|\e^2)\cr
(\e^2|\e^1) & (\e^2|\e^2)\cr}\right) =
   \left(\matrix{
m &0\cr
0 & 1\cr}\right) ,\; m\in \Nset,
\eeq
and let ${\bf Q}=\e^*_1$ (where $\{ \e^*_i \}$ stands for
the dual basis in $L^*$ satisfying $(\e^*_i|\e^j)=\d^i_j$ ).
Consider a $\Zset_M$ gauge group in $\A(L)$ generated by an
automorphism of type (\ref{3.7}) with
\beq\label{3.15}
\boldsymbol{\beta}=\boldsymbol{\beta_M}=
\frac{1}{M}(\mu \e^*_1+\e^*_2)\;\;\; 0<\mu<M.
\eeq
$\A(L)$ has $m$ (untwisted) representations with partition
functions
\beq\label{3.16}
\chi_l(\tau,\boldsymbol{\alpha})=K_l(\tau,\alpha^1;m)
K_0(\tau,\alpha^2;1) \;\;
\mathrm{for}\;\; \boldsymbol{\alpha}=\alpha^i\e^*_i,
\eeq
where $K_l$ are the 1-dimensional lattice characters
(see, e.g., \cite{pt})
\beq\label{3.17}
K_l(\tau,\alpha;m)=\frac{1}{\eta(\tau)}
\sum_{n\in\Zset} q^{\frac{m}{2}(n+\frac{l}{m})^2}
\ex^{2\pi i \alpha( n+\frac{l}{m})}
\eeq
$l$ mod $m$ (for $m$ even a canonical choice is
$1-\frac{m}{2}\leq l\leq \frac{m}{2}$).

The lift $b=\ex^{2\pi i H_0^{\boldsymbol{\beta}}} \
(H_0^{\boldsymbol{\beta}}=
\frac{1}{M}(\mu  H_0^{\e^*_1}+H_0^{\e^*_2}) )$ of the
automorphism group $\Zset_M$ to the $\A(L)$ modules $\H_l$ is
$\Zset_{mM}$. Setting
\beqa
&&\s^*(\ex^{2\pi i \boldsymbol{\alpha}})=
\ex^{-2\pi i \frac{\alpha}{m}\s}\;
(\mathrm{for} \; \boldsymbol{\alpha}=\alpha (\mu\e^*_1+\e^*_2) ),\nn
&&\boldsymbol{\beta}=\beta (\mu\e^*_1+\e^*_2); \;\
\ex^{2\pi i\frac{\alpha}{m}},\;
\ex^{2\pi i\frac{\beta}{m}}\in\Zset_{mM},\;\;
|\G_{\boldsymbol{\beta}}|=mM \nn
\eeqa
we can evaluate the finite sum in (\ref{3.11}) with the result
\beq\label{3.18}
\chi_{l,\s}^\beta(\tau,\z)=
\frac{1}{\eta^2(\tau)}\sum_{n_1,n_2\in\Zset}
\ex^{2\pi i(n_1 + \frac{l-\mu\beta}{m}) \z }
q^{\frac{m}{2}(n_1+\frac{l-\mu\beta}{m})^2+\h (n_2-\beta)}
\d_{m(\mu n_1+n_2)+\mu l}^{\s \ \mathrm{mod}\ mM}.
\eeq
It follows that $\s-\mu l=0$ mod $m$, in accord with
the first statement in Proposition 3.1, and we can take
the Kronecker $\d$ into account setting (for $M,\mu$
coprime)
\beqa\label{3.19}
&&n_1=\frac{\s-\mu l}{m} +r+M n,\;\;\;
n_2=\frac{\mu l -\s}{m}(\mu-1)-\mu r +m n',\nn
&&r=0,1,\ldots,M-1\;\;\;\;(n,n'\in\Zset).
\eeqa
Inserting in (\ref{3.18}) and using once more (\ref{3.17})
we find
\beqa\label{3.20}
\chi_{l,\s}^{\nu/M}(\tau,\z)=
\sum_{r=0}^{M-1} K_{M[\s-(\mu-1)l]
-\mu\nu+mMr}(\tau,M\z;mM^2)\times \nn
\times K_{M\frac{l\mu-\s}{m}(\mu-1)-\nu -\mu Mr}
(\tau,0;M^2)
\eeqa

{\bf Remark 3.2.} A non-analytic factor,
$\ex^{-\frac{\pi}{m} \frac{(Im\ \z)^2}{Im \ \tau}}$,
is to be assigned to the character in order to ensure its
$V:\z\rightarrow \z+\tau $ covariance (see \cite{cz}).

{\bf Remark 3.3.} Given the automorphism $\r$ there is
a ``gauge" freedom in the choice of $b$ and
$\boldsymbol{\beta}$ in
(\ref{3.7}) (in the example (\ref{3.15}) it amounts to
adding to $\boldsymbol{\beta}$ an arbitrary vector of
$L^*$). Theorem 4.2
of \cite{kt} guarantees that the resulting orbifold models
do not depend on this choice.

{\em To sum up:}  quantum field theory realization of $\W$ are
provided by U(1)$^{\times c}$ current algebras which also allow
to construct a large family of RCFT extensions of such algebras.
These leave room for additional physical requirements having in
mind specific applications, - e.g., to quantum Hall systems.

The present paper is an outgrow of previous work in
collaboration with Bojko Bakalov to whom we are indebted,
in particular, for an indication of the relevance of Schur's
polynomials in the present context. The work has been supported
in part by the Bulgarian Foundation for Scientific Research
under contract F-404.



\end{document}